\documentclass[a4paper,10pt]{article}
\usepackage{amsmath}
\newcommand{\hx}{{\hat x}}
\newcommand{\hPsi}{{\hat \Psi}}

\newcommand{\hPhi}{{\hat \Phi}}
\newcommand{\hPhid}{{{\hat \Phi}^{\dagger}}}
\newcommand{\hA}{{\hat A}}
\newcommand{\hD}{{\hat D}}
\newcommand{\hF}{{\hat F}}
\newcommand{\hU}{{\hat U}}
\newcommand{\m}{{\widetilde m}}
\newcommand{\Phid}{{\Phi^{\dagger}}}

\newcommand{\D}{{\mathcal D}}
\newcommand{\Du}{\overset{{\mbox{\tiny u}}}{\D}}
\newcommand{\Dbar}{{\overline{\mathcal D}}}
\newcommand{\n}{\nabla}
\newcommand{\nbar}{{\overline \nabla}}
\newcommand{\nablau}{\overset{{\mbox{\tiny u}}}{\nabla}}
\newcommand{\muz}{\mu_{o}}

\newcommand{\vt}{{\vec \theta}}

\newcommand{\tdB}{{{\vec \theta} \cdot {\vec B}}}
\newcommand{\tddB}{{{\vec \theta}\cdot \partial_{0}{\vec B}}}
\newcommand{\gz}{\overset{{\mbox{\tiny 0}}}{g}}
\newcommand{\go}{\overset{{\mbox{\tiny 1}}}{g}}

\newcommand{\Gammau}{\overset{{\mbox{\tiny u}}}{\Gamma}}
\newcommand{\gammau}{\overset{{\mbox{\tiny u}}}{\gamma}}
\newcommand{\None}{\overset{{\mbox{\tiny{1}}}}{N}}
\newcommand{\Ntwo}{\overset{{\mbox{\tiny{2}}}}{N}}
\begin{document}
\title{ $ U(1) $ gauge invariant noncommutative Schr\"odinger theory and gravity}
\author{B. Muthukumar\thanks{e-mail: muthu@theory.saha.ernet.in}\\
Saha Institute of Nuclear Physics,\\ 1/AF, Bidhan nagar,\\ Kolkata 700 064, India.}
\date{}
\maketitle

\begin{abstract}
We consider the complex, massive Klein-Gordon field living in the noncommutative space, and coupled to noncommutative electromagnetic fields. After employing the Seiberg-Witten map to first order, we analyze the noncommutative Klein-Gordon theory as $ c $, the velocity of light, goes to infinity. We show that the theory exhibits a regular "magnetic" limit only for certain forms of magnetic fields. The resulting theory is nothing but  the Schr\"odinger theory in a gravitational background generated by the gauge fields. 
\end{abstract}

\section{Introduction}
 
Gauge theories on noncommutative spaces, which arise as a limit of string theory, exhibit several interesting properties \cite{Seiberg,Douglas}. Noncommutative (NC) spaces are defined by the commutation relation $ \left[ \hx^{\mu}, \hx^{\nu} \right] = i \theta^{\mu\nu},  $ where $ \theta $ is a constant antisymmetric matrix. Such a commutator amounts to replace the ordinary products of functions in commutative coordinate space by the Moyal star product defined by
\begin{equation}
f(x)*g(x)= \left. \mathrm{exp}\left( \frac{i}{2} \theta^{\mu\nu}\partial_{\mu}^y \partial_{\nu}^z \right) f(y)g(z)\right|_{y=z=x}. \label{Star} 
\end{equation} 
 This product is noncommutative, associative and highly non-local. Local-gauge invariant physical observables in ordinary gauge theories, when pass over to noncommutative spaces, acquire this non-local nature. In fact, it was shown in \cite{Gross} that the translation of the gauge fields along the noncommutative directions is equivalent to gauge transformation plus a constant shift of the gauge field. To put it differently, a gauge transformation can generate a spatial translation and that is why we can not have  local gauge-invariant observables in NC gauge theories. This subtle property of NC gauge theories reveals their similarity to general relativity where translations are equivalent to gauge transformations, i.e., the general coordinate transformations.

Exploiting the novel mapping---called the Seiberg-Witten map---between the NC gauge fields and their commutative analogue  \cite{Seiberg},  the analogy to gravity was further corroborated in \cite{Rivelles}. It was shown that if ordinary fields are employed, then the analogy with gravity shows up in such a vivid manner that the effect of noncommutativity can be interpreted as a gravitational background that depends on the dynamical gauge fields. The analogy to gravitational coupling of the NC theory of scalar fields coupled to NC $ \hU(1) $ gauge fields was also clearly established in \cite{Rivelles}. The fields considered in \cite{Rivelles} are real/complex, massless Klein-Gordon fields coupled to NC $ \hU(1) $ gauge fields, and it was shown that after performing the Seiberg-Witten map, real and complex fields couple to different  gravitational backgrounds and that real field couples more strongly than the complex field. 

In this paper, we consider the nonrelativistic limit of the NC Klein-Gordon equation for a massive, complex field coupled to NC electromagnetic fields. This leads to the NC Schr\"odinger equation, which also  provides the quantum mechanical description of nonrelativistic particles. The quantum mechanics on NC spaces has been studied extensively in the literature because of its relevance in the one particle sector of NC field theories in the free field, or weak coupling case [5-20]. In the nonrelativistic sector, the noncommutative systems coupled to external electromagnetic fields have been dealt with, for example, in \cite{Chaichian,Nair,Bellucci,Kokado,Ho,Zhang,Chakraborty} but a detailed study in the spirit of \cite{Rivelles} is lacking. We provide here such an analysis.

The paper is organized as follows.

In Section 2, we discuss the massive NC $ \hU(1) $ gauge invariant NC Klein-Gordon theory along  
the lines of \cite{Rivelles}. We assume the noncommutativity only among the spatial coordinates; the temporal coordinate is taken to be commuting with the spatial ones. After employing the Seiberg-Witten map (SW map), the theory to first order in $ \theta  
$ is nothing but the conventional $ U(1) $ gauge invariant Klein-Gordon theory in a gravitational  
background generated by the gauge fields, and, the mass of the field acquires a field dependent first order  
correction.

Section 3  is a brief review of the Newton-Cartan spacetime structures that the nonrelativistic limit of the NC Klein-Gordon theory will have. We provide only the necessary details that will be used in the rest of the paper.

Finally, in Section 4 we derive the NC Schr\"odinger equation from the SW-mapped NC Klein-Gordon equation. We employ the method known in the literature as the high frequency phase factor method \cite{Schiff,Duval2}. As noted in \cite{Bellac} for the case of Galilean electromagnetism, two distinct nonrelativistic (NR) limits---called the {\em magnetic} limit and the {\em electric} limit---are possible when we have electromagnetic fields. We consider the magnetic limit.  We show that this limit exists only for specific forms of the magnetic field ($ {\vec B} $) that need to obey the condition that $ {\vec \theta} \cdot \partial_{\mu} {\vec B} $ should be  zero, where $ \theta^i $ is equal to $ (1/2) \epsilon^{ijk} \theta_{jk} $. The resulting nonrelativistic equation is the Schr\"odinger equation in a gravitational background generated by the gauge fields. This equation corresponds to the generally covariant Schr\"odinger equation (in a special gravitational gauge) that was originally obtained in \cite{Kuchar,Duval2}.  While the underlying geometrical structure  has the regular magnetic limit if and only if $ {\vec \theta} \cdot \partial_i {\vec B} $ is zero, the high frequency phase factor method gives consistent limit if and only if $ {\vec \theta} \cdot \partial_0 {\vec B} $ vanishes in addition to $ {\vec \theta} \cdot \partial_i {\vec B} $ being zero. We also compare this NR theory with the one obtained directly from the conventional $ U(1) $ gauge invariant Schr\"odinger theory deformed via the star product (\ref{Star}), and point out a few drawbacks in the later one.

\section{Noncommutative Klein-Gordon Theory}

As we are going to deal with the coupling of NC electromagnetic fields with the NC Klein-Gordon field, the NR limiting process needs to be handled with care. It was noticed in \cite{Bellac} for the case of conventional Galilean electromagnetism---the electromagnetic theory that is invariant under Galilean transformations---that when we have the electromagnetic fields, the limiting process will depend on the choice of the system of units and that choosing a system in which $ c $ enters in its very definition will lead to inconsistent results. In the case of conventional Maxwell theory, the way out is to use SI units which does not involve $ c $ in its definition but has it entered through the relation $ \epsilon_{o}\mu_{o} c^{2} = 1 $. But this relation suggests that we can not directly take the limit $ c \to \infty $, since both $ \epsilon_{o} $ and $ \mu_{o} $ are finite quantities. The limiting process suggested in \cite{Bellac} is to keep one of the quantities in the theory and eliminate the other by using the relation $ \epsilon_{o}\mu_{o} c^{2} = 1 $ before taking a limit (see also the end of Section 4). Therefore, one will land up with two different nonrelativistic limits: (i) the {\em magnetic} limit 
in which $ \mu_{o} $ is kept in the theory, and (ii) the {\em electric} limit in which $ \mu_{o} $ is eliminated. 

We make use of that technique and write the conventional theory in terms of $ \mu_{o} $.
When we deal with the NR limit of a relativistic theory, it is convenient to take the Minkowski metric to be $ \mathrm{diag}\left( \eta^{\mu\nu}\right) = \left( -1/c^2,1,1,1\right)  $  in the absence of electromagnetic fields (see, for example, \cite{Trautman}). In the presence of the electromagnetic fields,  
we take it to be 
\begin{eqnarray}
 \mathrm{diag}\left( \eta^{\mu\nu}\right) = \left( -1/( \mu_{o}c^2),\;1,\;1,\;1 \right), \label{Fmetric} 
\end{eqnarray}
since we consider the magnetic limit. That is, we employ the coordinate 4-vectors $ x^{\mu} = \left( t/\sqrt{\muz}, x^i\right)  $ and $ x_{\mu} = \left( -\sqrt{\muz}c^2 t, x^i\right)  $.  With these  
forms of covariant and contravariant vectors, the Maxwell tensor becomes
\begin{align}
F^{0i} = E^i/(\mu_{o}c^2),
\qquad F^{ij}= \epsilon^{ijk} B^k/\sqrt{\mu_{o}}\;,
 \label{ContraF} 
\end{align} 
and
\begin{eqnarray}
F_{0i} = - E^i/(\mu_{o}c^2),
\qquad F_{ij}= \epsilon^{ijk} B^k/\sqrt{\mu_{o}}\;.
\label{CoF} 
\end{eqnarray} 
The total action for the Klein-Gordon theory coupled to the ordinary $ U(1) $ gauge field is
\begin{eqnarray}
S_c = - \frac{1}{4} \int d^4 x \sqrt{-\eta}F^{\mu\nu} F_{\mu\nu} + \int d^4 x \sqrt{-\eta}\left[ \left( D^{\mu}\Phi\right)^{\dagger} \left(  D_{\mu}\Phi\right) + \frac{m^2c^2}{\hbar^2} \Phid\Phi \right], \label{S} 
\end{eqnarray} 
where $ \eta $ is the determinant of $ \eta_{\mu\nu} $, $ F_{\mu\nu} = \partial_{\mu} A_{\nu} - \partial_{\nu} A_{\mu} $ and $ D_{\mu}\Phi = \partial_{\mu}\Phi - ig A_{\mu}\Phi $; $ g $ is the gauge coupling constant.  Next, we use the Moyal star product (\ref{Star}) to get the noncommutative version of (\ref{S}). We assume that the noncommutativity exists only among the spatial coordinates, i.e., we take $ \theta^{0i} $ to be $ 0 $. We  follow the Ref. \cite{Rivelles} which treats the massless case. The action for the NC $ \hU(1) $ gauge field is given by
\begin{equation}
S_{A} = -\frac{1}{4}\int d^4 x \sqrt{-\eta} \hF^{\mu\nu} * \hF_{\mu\nu} \label{ShA},
\end{equation} 
where $ \hF_{\mu\nu}= \partial_{\mu}\hA_{\nu} -\partial_{\nu}\hA_{\mu} - ig \left( \hA_{\mu}*\hA_{\nu} - \hA_{\nu}*\hA_{\mu}\right) $. For the scalar field,
\begin{eqnarray}
S_{\Phi} = \int d^4 x \sqrt{-\eta}\left[ \left( \hD^{\mu}\hPhi\right)^{\dagger} *\left( \hD_{\mu}\hPhi \right) + \frac{m^2c^2}{\hbar^2} \hPhid * \hPhi \right], \label{ShatKG}
\end{eqnarray} 
where $ \hD_{\mu}\hPhi=\partial_{\mu}\hPhi- ig \hA_{\mu} * \hPhi$.  In this paper, we assume the scalar field to be complex. The real field has a different minimal coupling to the gauge field \cite{Rivelles}.

 The actions  (\ref{ShA}) and (\ref{ShatKG}) can be expressed in terms of the ordinary gauge fields $ A_{\mu} $ with $ F_{\mu\nu} =\partial_{\mu}A_{\mu} - \partial_{\nu}A_{\mu} $ and the conventional scalar field $ \Phi $ with $ D_{\mu}\Phi = \partial_{\mu}\Phi - ig A_{\mu}\Phi $ through the Seiberg-Witten map. To first order in $ \theta $, the map is \cite{Seiberg,Bichl,Rivelles} 
\begin{align}
\hA_{\mu} &= A_{\mu} - \frac{g}{2} \theta^{\alpha\beta} A_{\alpha}\left( \partial_{\beta} A_{\mu} + F_{\beta\mu}\right), \label{hA} \\
\hPhi &= \Phi - \frac{g}{2} \theta^{\alpha\beta} A_{\alpha} \partial_{\beta}\Phi. \label{hPhi} 
\end{align} 
Employing the map, the action  (\ref{ShA}) becomes
\begin{eqnarray}
S_A = \frac{-1}{4}\int d^4 x \sqrt{-\eta} F_{\mu}{}^{\rho} \left\lbrace  \eta^{\mu\nu} - g \left[  \theta^{\mu\gamma} F_{\gamma}{}^{\nu} + \theta^{\nu\gamma} F_{\gamma}{}^{\mu} + \frac{\eta^{\mu\nu}}{2}  \theta^{\gamma\sigma}F_{\gamma\sigma} \right] \right\rbrace F_{\nu\rho},
\end{eqnarray}
which, to first order in $ \theta $, can be written as
\begin{eqnarray}
S_A = -\frac{1}{4}\int d^4x \sqrt{-\eta}\; g^{\mu\nu} g^{\rho\sigma} F_{\mu\rho} F_{\nu\sigma}, \label{SA}
\end{eqnarray} 
where
\begin{eqnarray}
g^{\mu\nu} = \eta^{\mu\nu} - h^{\mu\nu} =\eta^{\mu\nu} - \left[ \frac{g}{2}\left( \theta^{\mu\gamma} F_{\gamma}{}^{\nu} + \theta^{\nu\gamma} F_{\gamma}{}^{\mu}\right) + \frac{g}{4} \eta^{\mu\nu} \theta^{\gamma\sigma}F_{\gamma\sigma} \right]. \label{ginverse} 
\end{eqnarray} 
Defining $ g_{\mu\nu} = \eta_{\mu\nu} + h_{\mu\nu} $ which to first order satisfies $ g_{\mu\lambda} g^{\lambda\nu} =\delta_{\mu}^{\nu} $, one can check that the determinant of $ g_{\mu\nu} $ is $ \eta $,  since the trace of $ h_{\mu\nu} $ would vanish. At this stage, one might think of (\ref{SA}) as the ordinary Maxwell theory coupled to a field dependent gravitational background. But the background cannot be treated as fixed since the fields themselves are dynamical \cite{Rivelles}. The NC Maxwell equations, i.e., the corresponding equations of motion for the action (\ref{SA}), are \cite{Guralnik}
\begin{alignat}{2}
\begin{aligned}
 \frac{\partial {\vec B}}{\partial t} + \n \times {\vec E} & = 0, & \qquad
\n \cdot {\vec B} & = 0,  \\
 \frac{\partial {\vec D}}{\partial t} - \n \times {\vec H} & =0,  & \qquad
\n \cdot {\vec D} &= 0, \label{Max} 
\end{aligned}
\end{alignat}
where
\begin{eqnarray}
\begin{aligned}
{\vec D} &= \frac{1}{\muz c^2} \left[ {\vec E} + \frac{e}{\hbar} \left(  - ({\vec \theta}\cdot {\vec B} ) 
{\vec E} + ({\vec \theta} \cdot {\vec E} ) {\vec B} + ( {\vec E} \cdot {\vec B}) {\vec \theta}\right) \right], \\
{\vec H} &= \frac{1}{\muz} \left[ {\vec B} + \frac{e}{\hbar} \left(  ({\vec \theta} \cdot {\vec B} ) {\vec B} - ( {\vec \theta} \cdot {\vec E} ) {\vec E} + \frac{1}{2} ( {\vec E}^2 - {\vec B}^2 ) {\vec \theta}\right) \right] \label{constitutive} 
\end{aligned}
\end{eqnarray}
are the constitutive relations.

On the other side, the action (\ref{ShatKG}) takes the form:
\begin{eqnarray}
S_{\Phi} = \int d^4 x \sqrt{-\eta}\left\lbrace g^{\mu\nu}  \left( D_{\mu} \Phi\right) ^{\dagger}  \left( D_{\nu}\Phi \right) + \frac{m^2 c^2}{\hbar^2} \left[ 1+\frac{g}{4} \theta^{\gamma\sigma} F_{\sigma\gamma} \right] \Phi^{\dagger} \Phi \right\rbrace . \label{SKG0} 
\end{eqnarray} 
Note that both (\ref{SA}) and (\ref{SKG0}) are invariant under ordinary $ U(1) $ gauge transformation. With the redefinition of mass:
\begin{eqnarray}
\m = m \left[ 1+\frac{g}{8} \theta^{\gamma\sigma} F_{\sigma\gamma} \right], \label{mbar} 
\end{eqnarray} 
the action (\ref{SKG0}) is written as
\begin{eqnarray}
S_{\Phi} = \int d^4 x \sqrt{-\eta}\left\lbrace g^{\mu\nu}  \left( D_{\mu} \Phi\right) ^{\dagger}  \left( D_{\nu}\Phi \right) + \frac{\m^2 c^2}{\hbar^2}  \Phi^{\dagger} \Phi \right\rbrace. \label{SKG} 
\end{eqnarray} 
This action is like that of the Klein-Gordon theory of the ordinary field $ \Phi $, with gauge field dependent mass, coupled to a gravitational background. Indeed, the Riemann tensor constructed out of $ g_{\mu\nu} $ does not vanish, and the Ricci tensor and the scalar curvature do vanish suggesting that the spacetime is not flat \cite{Rivelles}. We like to point out that it has been shown in \cite{Yang} that (\ref{ShA}) boils down to the form (\ref{SA}) to all orders in $ \theta $, but there are troubles to establish (\ref{SKG}) to all orders in $ \theta $ for the massless case\footnote{
We thank Prof. V.O. Rivelles for private communication drawing our attention to the reference \cite{Yang} and commenting on this regard.
}. For the massive case, the problem becomes more serious because in higher orders the mass term gives rise to derivatives of $ \Phi $. In this paper, we restrict up to the first order terms. 
The noncommutativity of spacetime not only causes a field dependent distortion of spacetime, which is elegantly interpreted using the mathematical tools of general relativity, but also provides the mass of the ordinary Klein-Gordon field a gauge field dependency.

We denote the Christoffel symbols by $ \{^{\kappa}_{\mu\nu}\}$, 
 and define the covariant derivative of a vector $ X^{\kappa} $ to be $ \nbar_{\mu} X^{\kappa} = \partial_{\mu} X^{\kappa} +  \{^{\kappa}_{\mu\nu}\} X^{\nu} $. Then the minimal coupling to the electromagnetic field in the curved spacetime may be defined by
$ \Dbar_{\mu}\Phi = \nbar_{\mu}\Phi - ig A_{\mu}\Phi $. The Klein-Gordon equation in curved spacetime, then, takes the form:
\begin{eqnarray}
g^{\mu\nu} \Dbar_{\mu} \Dbar_{\nu} \Phi - \frac{\m^2 c^2}{\hbar^2} \Phi = 0. \label{KGeqn} 
\end{eqnarray} 
We are interested in the NR limit of this equation. Since the energy involved in the NR limit is far less than $ mc^2 $, the term involving $ mc^2 $ is expected to decouple in the NR limit. In the case of the  conventional Klein-Gordon theory in flat spacetime, the limiting theory is achieved through the method of eliminating a high frequency phase factor $ e^{-imc^2t/\hbar} $ from the wave function \cite{Schiff}. In curved spacetime, employing a similar technique, the Klein-Gordon theory has been shown by Duval and K\"unzle \cite{Duval2} to reduce to the Newton-Cartan-Schr\"odinger equation \cite{Kuchar,Duval2}.   Newton-Cartan-Schr\"odinger equation is the generally spacetime covariant version of the Schr\"odinger equation first obtained by Kucha\v{r} \cite{Kuchar}, and later shown by Duval and K\"unzle \cite{Duval2}, using the language of fiber bundles, 
 that a prescription of minimal coupling of Newton-Cartan gravitational field to a complex scalar field leads to the same. The reference \cite{Christian} gives a coherent review of the Newton-Cartan theory of gravity and the generally covariant Schr\"odinger theory, and deals further with the quantum aspects of Newton-Cartan theory to show that the quantum field theory of Newtonian gravity coupled to Galilean relativistic matter is exactly soluble. A very brief review of the Newton-Cartan spacetime structures that the NR limiting theory of (\ref{KGeqn}) will have is provided in the next Section.

\section{The Newton-Cartan Spacetime Structures}

Geometrically speaking, the NR limit $ ( c \to \infty) $ can be viewed as a flattening or opening up 
of the light cones which would form the spacelike surfaces in the NR context. In the NR limit the spacetime structure is the general Galilean spacetime---a four-dimensional $ C^{\infty} $ manifold $ \mathcal{M} $  that is endowed with the three structures: (i) a positive, semi-definite space-metric $ \gamma^{\mu\nu} $ with signature $ (0\,+\,+\,+) $; (ii) a never vanishing covariant vector field $ T_{\mu} $ that corresponds to the time-metric $ T_{\mu\nu}=T_{\mu}T_{\nu} $ with signature $ (+\,0\,0\,0) $; (iii) a connection represented by the smooth covariant derivative $ \n_{\mu} $.  The following conditions need to be satisfied by these three structures \cite{Christian,Duval3}:
\begin{eqnarray}
\gamma^{\mu\nu} T_{\nu} = 0 ;\qquad \n_{\lambda} \gamma^{\mu\nu} = 0 ;\qquad \n_{\lambda}T_{\mu} = 0. 
\label{Gstructure} 
\end{eqnarray}  
The Galilean structure defined by these conditions is said to be integrable if
\begin{eqnarray}
\partial_{\left[ \lambda\right. } T_{\left. \mu \right] } = 0. \label{Gstructure2} 
\end{eqnarray} 
The integrability condition locally allows a time function $ T $ such that $ T_{\mu} = \partial_{\mu} T $ \cite{Kunzle,Christian}. Then the condition $ \n_{\lambda}T_{\mu} = 0 $ implies that $ T $ is determined up to a linear transformation $ T \to a T + b $, where $ a \; (> 0)  $ and $ b $ are constants \cite{Trautman}. 

Linear symmetric Galilean connection $ \Gamma^{\kappa}_{\mu\nu} $, such that $ \n_{\lambda} \gamma^{\mu\nu} = 0 $ and $ \n_{\lambda}T_{\mu} = 0 $, exists if and only if the structure $ (\gamma^{\mu\nu},\; T_{\mu}) $ is integrable. But the conditions (\ref{Gstructure}) and (\ref{Gstructure2}) determine the Galilean connection $ \Gamma^{\kappa}_{\mu\nu} $ only up to an antisymmetric tensor $ \mathcal{F}_{\mu\nu}$,  
i.e.,\footnote{We employ the symmetrization $ b^{(\mu\nu)} =  (b^{\mu\nu} + b^{\nu\mu})/2 $ and the antisymmetrization $ b^{[\mu\nu]} = ( b^{\mu\nu} - b^{\nu\mu} )/2 $.} \cite{Duval3}
\begin{eqnarray}
\Gamma^{\kappa}_{\mu\nu} = \Gammau {}^{\kappa}_{\mu\nu}
 + \gamma^{\lambda\kappa} \left[ T_{\left( \mu\right. } \mathcal{F}_{\left. \nu\right) \lambda}  \right], \label{Gconnection} 
\end{eqnarray} 
where $ \Gammau{}^{\kappa}_{\mu\nu} $ is given by
\begin{eqnarray}
\Gammau{}^{\kappa}_{\mu\nu} = \frac{1}{2} \gamma^{\lambda\kappa} \left[ \partial_{\mu}\gammau_{\nu\lambda} + \partial_{\nu}\gammau_{\lambda\mu} -\partial_{\lambda}\gammau_{\mu\nu} \right] + u^{\kappa} \partial_{\left( \nu\right. } T_{\left. \mu\right) } . 
\end{eqnarray} 
The vector $ u^{\mu} $ is  unit timelike,  i.e., $ u^{\mu} T_{\mu} = 1 $,  and it is interpreted as the four-velocity observer fields. A vector $ v^{\mu} $ is spacelike if $ \sqrt{v^{\mu} v^{\nu} T_{\mu\nu}} = 0 $, timelike if $ \sqrt{v^{\mu} v^{\nu} T_{\mu\nu}} > 0 $ and future-oriented if $ v^{\mu} T_{\mu} > 0 $ \cite{Christian}. The connection $ \Gammau{}^{\kappa}_{\mu\nu} $ is a unique symmetric Galilean connection such that the observer defined by $ u^{\mu} $ is geodesic and nonrotating,  i.e., 
\begin{align}
u^{\mu} \nablau_{\mu} u^{\nu}  = u^{\mu} &\left( \partial_{\mu} u^{\nu}  + \Gammau {}^{\nu}_{\mu\lambda} u^{\lambda} \right)  =0, \label{geodesic} \\ 
\gamma^{\mu\lambda}\nablau_{\lambda} u^{\nu} &= \gamma^{\nu\lambda} \nablau_{\lambda} u^{\mu}. \label{irrotational} 
\end{align}
The covariant metric $ \gammau_{\mu\nu} $, relative to the observer $ u^{\mu} $, is uniquely defined through the relations
\begin{eqnarray}
\gammau_{\mu\nu} u^{\nu} = 0, \qquad \gammau_{\mu\nu} \gamma^{\nu\lambda} = \delta^{\lambda}_{\mu} - u^{\lambda} T_{\mu}, \label{gammau-def} 
\end{eqnarray} 
and it is used to lower indices. 

The curvature tensor $ R^{\lambda}{}_{\mu\nu\alpha} $ given by
\begin{eqnarray}
R^{\lambda}{}_{\mu\nu\rho} = \frac{\partial\Gamma^{\lambda}_{\mu\nu}}{\partial x^{\rho}} - \frac{\partial\Gamma^{\lambda}_{\mu\rho}}{\partial x^{\nu}} + \Gamma^{\eta}_{\mu\nu} \Gamma^{\lambda}_{\rho\eta} - \Gamma^{\eta}_{\mu\rho} \Gamma^{\lambda}_{\nu\eta} \label{c-tensor} 
\end{eqnarray} 
has the following properties pertaining to the Galilean spacetime \cite{Duval3,Christian}
\begin{align}
\gamma^{\mu\left( \kappa\right. } R^{\left. \lambda\right) }{}_{\mu\nu\rho} = 0, \qquad T_{\lambda} R^{\lambda}{}_{\mu\nu\rho} = 0,
\qquad R^{\lambda}{}_{\mu(\nu\rho)}= 0, 
\qquad  R^{\lambda}{}_{[ \mu\nu\rho] }= 0 .
\end{align} 
The Ricci tensor and the scalar curvature are, respectively, $ R_{\mu\nu} = R^{\lambda}{}_{\mu\nu\lambda} $ and $ R = \gamma^{\mu\nu} R_{\mu\nu} $.

Newton-Cartan connections are  special cases of Galilean connections such that \cite{Duval3}
\begin{eqnarray}
R^{\lambda}{}_{\mu}{}^{\nu}{}_{\rho} = R^{\nu}{}_{\rho}{}^{\lambda}{}_{\mu}. \label{Curlfree} 
\end{eqnarray} 
Equivalently, the necessary and sufficient condition for a Galilean connection to be Newton-Cartan is that \cite{Duval3,Christian}
\begin{eqnarray}
\mathcal{F}_{\mu\nu} =  \partial_{\mu} {\mathcal A}_{\nu} - \partial_{\nu} {\mathcal A}_{\mu}, \label{mathcalF} 
\end{eqnarray} 
where $ A_{\mu} $ is interpreted as the combination of gravitational and inertial potentials with respect to the observer $ u^{\mu} $ \cite{Duval4}. The connection (\ref{Gconnection}) with the constraint (\ref{mathcalF}) is invariant under the following transformations:
\begin{eqnarray}
\begin{aligned}
u^{\mu} & \mapsto  u^{\mu} + \gamma^{\mu\nu} w_{\nu}, \\
A_{\mu} & \mapsto  A_{\mu} + \partial_{\mu} f + w_{\mu} - ( u^{\nu} w_{\nu} + \frac{1}{2} \gamma^{\nu\lambda} w_{\nu}w_{\lambda} ) T_{\mu} , \label{N-gtrans} 
\end{aligned}
\end{eqnarray} 
where $ f $ is a smooth function and $ w_{\mu} $ is a covariant vector which is defined only modulo $ T_{\mu} $. The transformations (\ref{N-gtrans}) are of the gauge group of the Newtonian gravitation theory \cite{Duval2,Ruede}. Note that the pair $ (u^{\mu}, {\mathcal A}_{\mu}) $ helps to uniquely determine the connection (\ref{Gconnection}), whereas the structure $ (\gamma^{\mu\nu}, T_{\mu}) $ alone does not. This suggests that we can consider the structure $ ( {\mathcal M}; \;\gamma^{\mu\nu}, T_{\mu}, u^{\nu}, {\mathcal A}_{\mu}) $ with the conditions (\ref{Gstructure}), (\ref{Gstructure2}) and (\ref{mathcalF}) instead of $ ({\mathcal M};\; \gamma^{\mu\nu}, T_{\mu}, \n_{\mu}) $ with (\ref{Gstructure}), (\ref{Gstructure2}) and (\ref{Curlfree}) to uniquely define the Newton-Cartan spacetime structures. 

%
%

\section{The Nonrelativistic Limit}

 We will be making use of the zeroth order in $ \theta $ of the equations (\ref{Max}) in their magnetic limit ($ \muz c^2 \to \infty $). To get them, we expand the fields $ {\vec E} $ and $ {\vec B} $ in orders of $ \theta $ as \cite{Berrino}
\begin{eqnarray}
{\vec E} = {\vec e} + {\vec E'} + {\vec E''} + \ldots, \qquad {\vec B} = {\vec b} + {\vec B'} + {\vec B''} + \ldots , \label{EBexpans} 
\end{eqnarray} 
substitute them in (\ref{Max}), and take the magnetic limit. The zeroth order equations of the resulting equations are
\begin{align}
 \frac{\partial {\vec b}}{\partial \tau} + \n \times {\vec e}  = 0,  \qquad
\n \cdot {\vec b}  = 0, \qquad
\n \times {\vec b}  =0,   \qquad
\n \cdot {\vec e} = 0, \label{Maxwell} 
\end{align}
where $ \tau $ is the absolute time in the nonrelativistic context. The equations (\ref{Maxwell}) are the same as the ones in \cite{Bellac}.

To discuss the NR limit form of a generally covariant relativistic theory, it is convenient to choose one fixed local coordinate system and the following one parameter $ (\lambda) $ family of Lorentz metrics \cite{Kunzle,Duval2}: 
\begin{align}
g^{\mu\nu} &= \gamma^{\mu\nu} + \lambda \go{}^{\mu\nu} + \mathcal{O}(\lambda^2), \label{Cmetric1}\\
g_{\mu\nu} &= - \left( \frac{1}{\lambda} \right)  T_{\mu} T_{\nu} + \gz{}_{\mu\nu} + {\mathcal O}(\lambda), \label{Cmetric2}
\end{align} 
 where
\begin{align}
\go{}^{\mu\nu} &= - u^{\mu} u^{\nu} + k^{\mu\nu}, \label{gone} \\
\gz{}_{\mu\nu} &= - 2 V T_{\mu} T_{\nu} + \gammau_{\mu\nu}, \label{gzero} 
\end{align} 
for a spacelike tensor $ k^{\mu\nu} $ and an arbitrary scalar $ V $. In the limit of flat metric, a comparison of (\ref{Cmetric2}) with (\ref{Fmetric}) tells that $ \lambda $ should be identified with $ (\muz c^2)^{-1} $. The vector $ u^{\mu} $ in Eqn.(\ref{gone}) is unit timelike \cite{Kunzle} and it is of special significance: Quoting \cite{Kunzle}, "Roughly speaking it gives the local time axis of the reference frames with respect to which one expands the relativistic theories in powers of $ \lambda $ from their Newtonian limit".  In the NR limit, the expansion of the Christoffel symbols become
\begin{eqnarray}
\begin{aligned}
 \{^{\kappa}_{\mu\nu}\} & = - \frac{2}{\lambda} \gamma^{\rho\kappa} T_{\left( \nu\right.} \partial_{\left. \left[ \mu\right) \right. } T_{\left. \rho \right] }  + \Gamma{}^{\kappa}_{\mu\nu}  - ( 4 V \gamma^{\rho\kappa}  + 2 \go{}^{\rho\kappa} )  T_{\left( \nu\right.} \partial_{\left. \left[ \mu\right) \right. } T_{\left. \rho \right] }
  + {\mathcal O}(\lambda) ,
\end{aligned}
\end{eqnarray} 
where $ \Gamma{}^{\kappa}_{\mu\nu} $ is given by
\begin{eqnarray}
\Gamma{}^{\kappa}_{\mu\nu} = \frac{1}{2} \gamma^{\lambda\kappa} \left[ \partial_{\mu}\gammau_{\nu\lambda} + \partial_{\nu}\gammau_{\lambda\mu} -\partial_{\lambda}\gammau_{\mu\nu} \right] + u^{\kappa} \partial_{\left( \nu\right. } T_{\left. \mu\right) } + \gamma^{\kappa\rho} \partial_{\rho} V T_{\mu} T_{\nu} . \label{Limit-connection} 
\end{eqnarray} 
Note that the connection $ \{^{\kappa}_{\mu\nu}\} $ constructed out of the metric $ g_{\mu\nu} $ has a regular limit as $ \lambda \to 0 $ if and only if $ \partial_{\left[ \lambda \right.} T_{\left. \mu \right] } = 0 $, i.e., if the Galilean structure is integrable---a theorem proved in \cite{Kunzle}. Therefore, in this case the connection $ \{^{\kappa}_{\mu\nu}\} $ takes the form:
\begin{eqnarray}
\begin{aligned}
 \{^{\kappa}_{\mu\nu}\} & =  \Gamma{}^{\kappa}_{\mu\nu} + {\mathcal O}(\lambda), \label{Lconnection} 
\end{aligned}
\end{eqnarray} 
where $ \Gamma{}^{\kappa}_{\mu\nu} $, given by (\ref{Limit-connection}), is the particular case of (\ref{Gconnection}) for which $ {\mathcal A}_{\mu} = - V T_{\mu} $.

In the case of the noncommutative Klein-Gordon theory described by (\ref{SKG}), the induced background gravitational metric is 
\begin{eqnarray}
g_{\mu\nu} = \eta_{\mu\nu} + \left[ \frac{g}{2}\left( \theta_{\mu\gamma} F^{\gamma}_{\;\;\nu} + \theta_{\nu\gamma} F^{\gamma}_{\;\;\mu}\right) + \frac{g}{4} \eta_{\mu\nu} \theta^{\gamma\sigma}F_{\gamma\sigma} \right], \label{g} 
\end{eqnarray} 
and the inverse metric is given by (\ref{ginverse}). With the help of (\ref{ContraF}) and (\ref{CoF}), one can show the components of the metric to be
\begin{align}
\begin{split}
g_{00} &= -\muz c^2 ( 1+\frac{g}{2\sqrt{\muz}}  \tdB ), \qquad
g_{0i} =  \frac{g}{2} ( {\vec E} \times \vt )_i, \\
g_{ij} &= \delta_{ij} ( 1 - \frac{g}{2\sqrt{\muz}} \tdB ) + \frac{g}{2\sqrt{\muz}} \left(  \theta^i B^j + \theta^j B^i \right), \label{g-comp} 
\end{split}
\end{align}
and that of the inverse metric to be
\begin{align}
\begin{split}
g^{00} &= - \frac{1}{\muz c^2} ( 1 - \frac{g}{2\sqrt{\muz}} \tdB ), \qquad
g^{0i} = \left( \frac{g}{2}\right)  \left( \frac{1}{\muz c^2} \right) ( {\vec E} \times {\vec \theta})_i, \\ 
g^{ij} &= \delta_{ij} ( 1 + \frac{g}{2\sqrt{\muz}} \tdB ) - \frac{g}{2\sqrt{\muz}} \left(  \theta^i B^j + \theta^j B^i \right). \label{ginv-comp} 
\end{split}
\end{align} 

A comparison of the above components with that of (\ref{Cmetric1}) and (\ref{Cmetric2}) yields, to first order in $ \theta $, the following explicit forms for $  T_{\mu}, u^{\mu}, \gamma^{\mu\nu}, \gammau_{\mu\nu}, V $ and $ k^{\mu\nu} $:
\begin{align}
& T_{\mu} = \left( ( 1 + \frac{g\tdB}{4\sqrt{\muz}} ),\;0,\;0,\;0\right); \label{T-comp} \\
& u^{\mu} = \left( ( 1- \frac{g\tdB}{4\sqrt{\muz}} ) ,\; \frac{-g}{2}( {\vec E} \times \vt )_i \right); \label{u-comp} \\
\begin{split}
& \gamma^{0\mu} = 0,\;\;\;\; \gamma^{ij} =( 1+ \frac{g\tdB}{2\sqrt{\muz}} ) \delta_{ij} -  \frac{g}{2\sqrt{\muz}}\left( \theta^i B^j + \theta^j B^i\right); \label{gamma-comp} 
\end{split}\\
\begin{split}
& \gammau_{00} = 0 , \;\;\;\; \gammau_{0i} = \frac{g}{2}( {\vec E} \times \vt )_i, \;\;\;\;
 \gammau_{ij} = ( 1 - \frac{g\tdB}{2\sqrt{\muz}} ) \delta_{ij} +  \frac{g}{2\sqrt{\muz}}\left( \theta^i B^j + \theta^j B^i\right). \label{gammau-comp} 
\end{split}
\end{align} 

The equation (\ref{gzero}), together with the definition that $ \gammau_{\mu\nu} u^{\nu} = 0 $, leads to $ V = 0 $. Therefore,
\begin{eqnarray}
{\mathcal A}_{\mu} = - V T_{\mu} = 0. \label{A-comp}
\end{eqnarray} 

 The property that $ k^{\mu\nu} $ is a spacelike tensor, i.e., $ k^{\mu\nu} T_{\nu} = 0 $, together with $ T_0 \neq 0 $ and $ T_i = 0 $, determines that $ k^{\mu 0} = 0 $. Moreover, a comparison of the expression (\ref{gone}) with the $ {\mathcal O}(\lambda) $ term of (\ref{ginv-comp}) gives that $ k^{ij} =0 $. Therefore, $ k^{\mu\nu}  $ is also equal to zero. 

The explicit forms of the components of the connection constructed from the metric components (\ref{g-comp}) and (\ref{ginv-comp}) are:
\begin{align}
\{^{0}_{00}\} &= \frac{g}{4\sqrt{\muz}} \tddB + {\mathcal O}(\lambda) + {\mathcal O}(\theta^2),\\
\{^{0}_{\mu j}\} &=  {\mathcal O}(\lambda) + {\mathcal O}(\theta^2), \\
\{^{i}_{00}\} &=  \frac{1}{2} g^{ij} \left( \frac{1}{\lambda}\right) \frac{g}{2\sqrt{\muz}} ({\vec \theta} \cdot \partial_j {\vec B}) + \frac{g}{2} (\partial_0 {\vec E} \times {\vec \theta} )_i + {\mathcal O}(\theta^2), \label{divergent1} \\
\{^{i}_{kl}\} &= \frac{g}{4\sqrt{\muz}} \left[ \theta^l \partial_k B^i + \theta^i \partial_k B^l + \theta^i \partial_l B^k + \theta^k \partial_l B^i \right. 
\nonumber \\  
               &\phantom{=} \left. - \theta^k \partial_i B^l - \theta^l \partial_i B^k \right]  +{\mathcal O}(\lambda) + {\mathcal O}(\theta^2),\\
\{^{k}_{0j}\} &= \frac{g}{4\sqrt{\muz}} \left[ - (\tddB) \delta_{jk} + \theta^j \partial_0 B^k + \theta^k \partial_0 B^j \right]  \nonumber \\
 &\phantom{=} + \frac{g}{4}\left[ ( \partial_j {\vec E} \times {\vec \theta} )_k - ( \partial_k {\vec E} \times {\vec \theta} )_j  \right] + {\mathcal O}(\theta^2).
\end{align} 

We like to draw the reader's attention to the Eqn.(\ref{divergent1}) where $ \lambda $ appears in the denominator, and hence the component $ \{^{i}_{00}\} $ diverges in the limit $ \lambda \to 0 $. It has a regular limit if and only if  the factor $ ({\vec \theta} \cdot \partial_j {\vec B}) $ becomes zero. This is also compatible with the theorem previously stated. That is, the integrability condition $ \partial_{\lambda} T_{\mu} = \partial_{\mu} T_{\lambda} $ gives that
\begin{eqnarray}
{\vec \theta} \cdot \partial_j {\vec B} = 0. \label{conditioned-B}
\end{eqnarray} 
This condition, being of first order in $ \theta $, is peculiar to the noncommutative theory. In the limit $ \theta \to 0 $, we do not have any such restriction on the form of $ {\vec B} $ in the NR limit (see  \cite{Schiff}).

In the limit $ \lambda \to 0 $, the components of the Galilean connection (\ref{Limit-connection}) are, then, given by
\begin{align}
\Gamma{}^{0}_{00} & = \frac{g}{4\sqrt{\muz}} \tddB ; \qquad \Gamma{}^{0}_{\mu j} = 0; \qquad  
\Gamma{}^{i}_{00} = \frac{g}{2} (\partial_0 {\vec E} \times {\vec \theta} )_i; \nonumber \\
\Gamma{}^{i}_{kl} & = \frac{g}{4\sqrt{\muz}} \left[ \theta^l \partial_k B^i + \theta^i \partial_k B^l + \theta^i \partial_l B^k + \theta^k \partial_l B^i - \theta^k \partial_i B^l - \theta^l \partial_i B^k \right] ;\label{Gammau-comp} \\
\Gamma{}^{k}_{0j} & = \frac{g}{4\sqrt{\muz}} \left[ - (\tddB) \delta_{jk} + \theta^j \partial_0 B^k + \theta^k \partial_0 B^j \right] + \frac{g}{4}\left[ ( \partial_j {\vec E} \times {\vec \theta} )_k - ( \partial_k {\vec E} \times {\vec \theta} )_j  \right]. \nonumber
\end{align}

Next, we classify $ \tdB $ that obey the condition (\ref{conditioned-B}) into two cases: (i)  $ \tddB $ is zero and (ii)  $ \tddB $ is not zero. For the case (i), the mass (\ref{mbar}) becomes a constant; however, for the case (ii), it becomes a function of time. In the following, we consider only the case of constant mass, and, at the end of this Section, we comment on the other case in which the mass is a function of time. Note that the components of the connection are of first order. Therefore, making use of (\ref{EBexpans}) and the equations (\ref{Maxwell}), the components of the curvature tensor (\ref{c-tensor}) are worked out to be:
\begin{eqnarray}
\begin{aligned}
R^{j}{}_{00l} & = - \frac{g}{4\sqrt{\muz}} \left[ \theta^l \partial_0\partial_0 b^j + \theta^j \partial_0\partial_0 b^l\right] + \frac{g}{4}\left[ (\partial_0 \partial_l {\vec e} \times {\vec \theta})_j + (\partial_0 \partial_j {\vec e} \times {\vec \theta})_l \right] ; \\
R^{i}{}_{j0l} &= \frac{g}{4\sqrt{\muz}} \left[ \theta^i \partial_j \partial_0 b^l - \theta^j \partial_i \partial_0 b^l \right] 
  + \frac{g}{4} \left[ (\partial_l \partial_j {\vec e} \times {\vec \theta} )_i - (\partial_l \partial_i {\vec e} \times {\vec \theta} )_j \right]. \label{R-comp} 
\end{aligned} 
\end{eqnarray}
Other components which are not symmetric to (\ref{R-comp}) are zero. Moreover, the Ricci tensor and the scalar curvature vanish. The nonvanishing of the curvature tensor suggests that the spacetime is not flat. Note that when both $ {\vec e} $ and $ {\vec b} $ are constant, the spacetime will not be curved as expected from (\ref{g}). Incidentally, a rigid Galilean observer frame defined by, say $ v^{\mu} $, needs to satisfy $ \gamma^{\alpha\beta} \n_{\beta} v^{\mu} = - \gamma^{\alpha\mu} \n_{\beta} v^{\beta} $ \cite{Kuchar}. But one can show that the observer frame described by (\ref{u-comp}) does not satisfy this requirement suggesting that the frame is not a rigid one.

So far we considered the underlying spacetime structure in the NR limit. To achieve the NR limit of (\ref{KGeqn}), following \cite{Schiff,Duval2} we write the field $ \Phi $ and the time coordinate $ x^0 $ as
\begin{align}
\Phi &= e^{ - \frac{i\alpha }{\lambda}x^0} \Psi, \label{hffp} \\
x^0 &= T + \lambda \None + \lambda^2 \Ntwo + {\mathcal O}(\lambda^3), \label{xzero}
\end{align} 
where $ \None $ and $ \Ntwo $ are real scalars, and $ \alpha$ is equal to $ \m/(\hbar\sqrt{\muz}) $. 
The zeroth order of $ x^0 $, i.e., $ T $ may be chosen to be equal to $ ( 1 + \frac{g\tdB}{4\sqrt{\muz}} ) (\tau/\sqrt{\muz}) $. One can also replace  $ \Psi $ in (\ref{hffp}) with $ ( 1 + \lambda \stackrel{{\mbox{\tiny{1}}}}{\omega} + {\mathcal O}(\lambda^2) ) \Psi $, $ \stackrel{{\mbox{\tiny{1}}}}{\omega} $ being some real scalar, as used in \cite{Duval2}. As we are interested only in the zeroth order in $ \lambda $ of (\ref{KGeqn}), the form (\ref{hffp}) suffices our purpose. We write the Klein-Gordon equation (\ref{KGeqn}) as
\begin{eqnarray}
 {\mathcal E}_{{\mbox{\tiny KG}}}[\Phi] = g^{\mu\nu} \Dbar_{\mu} \Dbar_{\nu} \Phi - \frac{\alpha^2}{\lambda} \Phi = 0.
\end{eqnarray} 
After making use of (\ref{Cmetric1}), (\ref{Lconnection}), (\ref{hffp}) and (\ref{xzero}), $ {\mathcal E}_{{\mbox{\tiny KG}}} $ becomes
\begin{eqnarray}
\begin{aligned}
{\mathcal E}_{{\mbox{\tiny KG}}}[\Phi] & =  e^{ - \frac{i\alpha }{\lambda}x^0}\left[ \left( \frac{2\alpha}{\hbar}\right)  {\mathcal E} [\Psi] - 2 i \alpha \gamma^{\mu\nu} (\partial_{\mu}\None)    (\Du_{\nu} \Psi)  +   2\alpha^2 u^{\mu} (\partial_{\mu}\None) \Psi \right. \\
& \phantom{=} \left.  +   \alpha^2 \gamma^{\mu\nu} (\partial_{\mu}\None) 
 (\partial_{\nu}\None) \Psi  - i\alpha \gamma^{\mu\nu} (\nablau_{\mu}\partial_{\nu}\None)  \Psi + {\mathcal O}(\lambda)  \right] = 0, \label{KGLimit}
\end{aligned} 
\end{eqnarray} 
where $ \Du_{\mu} \Psi = \nablau_{\mu} \Psi - ig A_{\mu} \Psi $ and
\begin{eqnarray}
{\mathcal E}[\Psi] = \frac{\hbar}{2\alpha} \gamma^{\mu\nu} \Du_{\mu} \Du_{\nu} \Psi + i\hbar u^{\mu} (\Du_{\mu} \Psi) + \frac{i\hbar}{2} (\nablau_{\mu} u^{\mu}) \Psi .
\end{eqnarray} 
Moreover, it was shown in \cite{Duval2} that in the limit $ c \to \infty $, $ \None $ needs to be a constant in which case (\ref{KGLimit}) reduces to
\begin{eqnarray}
{\mathcal E}_{{\mbox{\tiny KG}}}[\Phi] = e^{ - \frac{i\alpha }{\lambda}x^0}\left[ \left( \frac{2\alpha}{\hbar}\right) {\mathcal E} [\Psi] + {\mathcal O}(\lambda)  \right] = 0 .
\end{eqnarray} 
Comparing the terms of like powers of $ \lambda $, the zeroth order equation is given by
\begin{eqnarray}
{\mathcal E}[\Psi] = \frac{\hbar}{2\alpha} \gamma^{\mu\nu} \Du_{\mu} \Du_{\nu} \Psi + i\hbar u^{\mu} (\Du_{\mu} \Psi) + \frac{i\hbar}{2} (\nablau_{\mu} u^{\mu}) \Psi
 = 0.
\end{eqnarray} 
Keeping in mind that $ \Gammau=\Gamma $ and that $ \tddB = 0 $, $ \nablau_{\mu} u^{\mu} $ in the last term can be shown to be zero.  This, finally,  leads to the Schr\"odinger equation in the spacetime defined by the Galilean structure (\ref{T-comp}), (\ref{u-comp}), (\ref{gamma-comp}) and (\ref{A-comp}):
\begin{eqnarray}
{\mathcal E}[\Psi] = \frac{\hbar}{2\alpha} \gamma^{\mu\nu} \Du_{\mu} \Du_{\nu} \Psi + i\hbar u^{\mu} (\Du_{\mu} \Psi)
 = 0. \label{NCSeqn0} 
\end{eqnarray} 

When we go from the coordinate structure $ (\frac{\tau}{\sqrt{\muz}}, x^i) $ to $ (\tau, x^i) $, $ u^{\mu} $ goes to $ \sqrt{\muz} u^{\mu} $ as it is the four-velocity observer field. In this case, the equation (\ref{NCSeqn0}) becomes
\begin{eqnarray}
 \frac{\hbar^2}{2\m} \gamma^{\mu\nu} \Du_{\mu} \Du_{\nu} \Psi + i\hbar u^{\mu} (\Du_{\mu} \Psi)  = 0.  \label{NCSeqn1} 
\end{eqnarray} 
This equation corresponds to the generally covariant Schr\"odinger equation 
\begin{eqnarray}
\frac{\hbar^2}{2\m} \gamma^{\mu\nu} { \D}_{\mu} { \D}_{\nu} \Psi + i\hbar u^{\mu} ({ \D}_{\mu} \Psi) + \frac{i\hbar}{2} (\n_{\mu} u^{\mu}) \Psi 
 = 0 \label{NCSeqn} 
\end{eqnarray} 
obtained by Kucha\v{r} \cite{Kuchar}, and Duval and K\"unzle \cite{Duval2} (in a special gauge given by (\ref{u-comp}) and (\ref{A-comp}) ). The covariant derivative $ { \D}_{\mu} \Psi $ in (\ref{NCSeqn})
is defined by $ {\D}_{\mu}\Psi=\n_{\mu}\Psi-\frac{i\m}{\hbar}{\mathcal A}_{\mu}\Psi-igA_{\mu}\Psi $.

One can also show that the Lagrangian in (\ref{SKG}) goes over to the Lagrangian for the Schr\"odinger theory in curved spacetime as $ \lambda $ tends to zero (see \cite{Duval2} which deals with the general Lagrangian):
\begin{eqnarray*}
\left\lbrace g^{\mu\nu}  \left( D_{\mu} \Phi\right) ^{\dagger}  \left( D_{\nu}\Phi \right) + \frac{\m^2 c^2}{\hbar^2}  \Phi^{\dagger} \Phi \right\rbrace = \frac{2\m}{\hbar^2} \left( {\mathcal L}[\Psi] \right) + {\mathcal O}(\lambda),
\end{eqnarray*} 
where 
\begin{eqnarray}
{\mathcal L}[\Psi] = \frac{\hbar^2}{2\m} \gamma^{\mu\nu} \left( D_{\mu} \Psi\right) ^{\dagger}  \left( D_{\nu}\Psi \right) + \frac{i\hbar}{2} u^{\mu} \left(  \left( D_{\mu} \Psi\right) ^{\dagger} \Psi - \Psi^{\dagger} \left( D_{\mu}\Psi \right) \right). \label{NCS-Lag} 
\end{eqnarray} 

For the sake of completeness, we like to compare this Lagrangian with the one obtained directly from the Lagrangian for the conventional $ U(1) $ gauge invariant Schr\"odinger theory deformed by the star product (\ref{Star}). 

The action for the conventional Schr\"odinger field coupled to a $ U(1) $ gauge field is
\begin{eqnarray*}
S_c = \int d^4 x \left[ \frac{\hbar^2}{2 m} \left( D_{i} \Psi\right)^{\dagger}  \left( D_{i}\Psi \right) + \frac{i\hbar}{2} \left[  \left( D_{0} \Psi\right)^{\dagger} \Psi - \Psi^{\dagger} \left( D_{0}\Psi \right) \right] \right] .
\end{eqnarray*} 
The NC version is obtained through the star product (\ref{Star}):
\begin{eqnarray*}
S[\Psi] = \int d^4 x \left[ \frac{\hbar^2}{2 m} ( \hD_{i} \hPsi)^{\dagger} * ( \hD_{i}\hPsi) + \frac{i\hbar}{2} \left[  ( \hD_{0} \hPsi)^{\dagger} * \hPsi - \hPsi^{\dagger} * ( \hD_{0}\hPsi) \right] \right] .
\end{eqnarray*} 
After making use of the Seiberg-Witten map, the above action boils down to the form:
\begin{eqnarray}
\begin{aligned}
 S[\Psi] & = \int d^4 x \left[ \dfrac{\hbar^2}{2 m} \gamma^{ij} \left( D_{i} \Psi\right)^{\dagger}  \left( D_{j}\Psi \right)  + \frac{i\hbar}{2} u^i \left[  \left( D_{i} \Psi\right)^{\dagger} \Psi - \Psi^{\dagger} \left( D_{i}\Psi \right) \right] \right. \\
 & \phantom{=} + \left.  \dfrac{i\hbar}{2} \left( 1 - \dfrac{g\tdB}{2\sqrt{\muz}}\right) \left[  \left( D_{0} \Psi\right)^{\dagger} \Psi - \Psi^{\dagger} \left( D_{0}\Psi \right) \right] \right], \label{Sch-to-Sch} 
\end{aligned}
\end{eqnarray}
where the forms of  $ u^i $ and $ \gamma^{ij} $ are the same as in (\ref{u-comp}) and (\ref{gamma-comp}), but with one major difference: the magnetic field is {\em not} restricted by any condition like $ {\vec \theta} \cdot \partial_{\mu} {\vec B} $ should be zero. For the magnetic fields satisfying this condition, the Lagrangian in (\ref{Sch-to-Sch}) can be shown to be equivalent to (\ref{NCS-Lag})---up to an overall constant factor. Therefore, one might be tempted to start from (\ref{Sch-to-Sch}) and work towards giving an interpretation of the theory being embedded in a gravitational background generated by the gauge fields. But note that such an interpretation, especially when it comes to Eqn.(\ref{Gstructure2}), will be {\em ad hoc} and will not be as demanding as it happens in the derivation of the NR theory from the relativistic theory.  For other forms of the fields, not only the theory lacks an interpretation of various terms in the Lagrangian, but also it is not furnished with details about which NR limit it corresponds to. Or rather, the theory being nonrelativistic does not clearly distinguish between the two NR limits.  

Finally, we comment on the case of the time dependent mass in the magnetic limit, and on the electric limit.

In the case of time dependent mass, if we press on with the high frequency phase factor method, then we would end up with terms of order $ 1/\lambda $, which diverge in the limit $ \lambda \to 0 $. It is precisely for this reason we avoided taking $ \tdB $ to be a function of $ t $ which would  have otherwise  given an inconsistent limit. To put it in other words (keeping in mind the condition (\ref{conditioned-B})), for a consistent NR magnetic limit to exist, $ {\vec \theta} \cdot \partial_{\mu} {\vec B} $ should be zero.

Our consideration of the magnetic limit was based on the existence of the corresponding smooth limit of the   Maxwell equations (\ref{Max}). In the case of conventional Maxwell theory, the mere elimination of $ \muz $ by using the relation $ \epsilon_{o}\muz c^2 = 1 $ does not lead to a consistent and correct electric limit. A method has been prescribed in \cite{Bellac} to obtain the theory in the electric limit directly from the Maxwell equations. To quote \cite{Bellac}, "To obtain the {\em electric} limit, express the relativistic theory in terms of $ {\mathbf E} $ and a redefined magnetic field $ {\widetilde{\mathbf B}} = c^2 {\mathbf B} $, keeping $ \epsilon_o $ but eliminating $ \muz $ (written $ 1/\epsilon_o c^2 $), then let $ c \to \infty $". This prescription is based on the {\em a priori} knowledge of the limiting theory formulated using physical and group theoretical arguments in \cite{Bellac} itself. To understand whether such a prescription will be applicable in the case of the noncommutative theory, one needs to construct a consistent electric limit form of the NC electrodynamics. Perhaps a group theoretical approach along the lines of \cite{Bellac,Kunzle} might provide a better setting in the case of the electric limit form of the theory. We will report our progress elsewhere.

\section*{Acknowledgments}

It is a pleasure to thank Prof. P. Mitra for a careful reading of the manuscript and for his questions and  comments on it. Correspondence with Prof. V.O. Rivelles is gratefully acknowledged.

\end{document}